\documentclass[prl,twocolumn,showpacs,superscriptaddress]{revtex4}
\usepackage{graphicx}

\newcommand{\be}{\begin{equation}}
\newcommand{\ee}{\end{equation}}
\newcommand{\bea}{\begin{eqnarray}}
\newcommand{\eea}{\end{eqnarray}}

\newcommand{\f}{\frac}

\newcommand{\clockwise}{
	\begin{picture}(14,20)(-2,2)
	\thicklines
	\put(0,0){\framebox(10,10){{\it p}}}
	\put(4,-0.5){\vector(1,0){5}}
	\put(10.5,4){\vector(0,1){5}}
	\put(6,10.5){\vector(-1,0){5}}
	\put(-0.5,6){\vector(0,-1){5}}
	\end{picture}
	}
\newcommand{\counterclockwise}{
	\begin{picture}(14,20)(-2,2)
	\thicklines
	\put(0,0){\framebox(10,10){{\it p}}}
	\put(6,-0.5){\vector(-1,0){5}}
	\put(10.5,6){\vector(0,-1){5}}
	\put(4,10.5){\vector(1,0){5}}
	\put(-0.5,4){\vector(0,1){5}}
	\end{picture}
	}
\newcommand{\lal}{\put(0,5.8){\vector(-1,0){0}}} 
\newcommand{\lar}{\put(5,5.8){\vector(1,0){0}}}  
\newcommand{\ral}{\put(7,5.8){\vector(-1,0){0}}} 
\newcommand{\rar}{\put(12,5.8){\vector(1,0){0}}} 

\newcommand{\uau}{\put(6.1,12){\vector(0,1){0}}} 
\newcommand{\uad}{\put(6.1,7){\vector(0,-1){0}}} 
\newcommand{\dau}{\put(6.1,5){\vector(0,1){0}}} 
\newcommand{\dad}{\put(6.1,0){\vector(0,-1){0}}} 

\newcommand{\cross}{ \put(0,6){\line(1,0){12}} \put(6,0){\line(0,1){12}}}
\newcommand{\bep}{\begin{picture}(12,10)(0,2)}
\newcommand{\eep}{\end{picture}}

\newcommand{\aonevertex}{\bep \cross \lar \dau \rar \uau \eep}
\newcommand{\atwovertex}{\bep \cross \lal \dad \ral \uad \eep}
\newcommand{\bonevertex}{\bep \cross \lar \dad \rar \uad \eep}
\newcommand{\btwovertex}{\bep \cross \lal \dau \ral \uau \eep}
\newcommand{\conevertex}{\bep \cross \lar \dad \ral \uau \eep}
\newcommand{\ctwovertex}{\bep \cross \lal \dau \rar \uad \eep}

\newcommand{\DDW}{{\cal O}_{\rm DDW}}
\newcommand{\nf}{n_{\rm f}}
\newcommand{\nsf}{n_{\rm sf}}
\newcommand{\nsftwo}{n_{\rm sf}^2}

\begin{document}

\title{Resonating plaquette phase of a quantum six-vertex model}

\author{Olav  F. Sylju{\aa}sen}

\affiliation{NORDITA, Blegdamsvej 17, DK-2100 Copenhagen {\O}, Denmark}
\author{Sudip Chakravarty}
\affiliation{Department of Physics and Astronomy, University of California Los Angeles, Los Angeles, California 90095}

\date{\today}

\pacs{74.20.-z, 71.10.Hf, 74.10.+v}

\begin{abstract}
The simplest quantum generalization of the six-vertex model describes fluctuations of the order parameter of the $d$-density wave (DDW), believed to compete with superconductivity in the high-$\mathrm{T_{c}}$ superconductors. The ground state of this model undergoes a first order transition from the DDW phase to a resonating plaquette phase as the quantum fluctuations are increased, which is explored with the help of quantum Monte Carlo simulations and analytic considerations involving the $n$-vector ($n=2$) model with cubic anisotropy. In addition to finding a new quantum state, we show that the DDW is robust against a class of quantum fluctuations of its order parameter. The inferred finite temperature phase diagram contains unsuspected multicritical points. 
\end{abstract}

\maketitle

The vertex models have unusual building blocks: arrows joined at a site forming the vertex. Nonetheless the statistical mechanics of these models are well defined, elegant, and have unexpected connections to more intuitively familiar models. For example, the transfer matrix of the classical six-vertex model is the XXZ-Hamiltonian \cite{Lieb} of Heisenberg spins and the corresponding transfer matrix of the eight vertex model is the XYZ-Hamiltonian.~\cite{Baxter} Both of these models are solved by powerful mathematical methods with far reaching consequences. These vertex models are not simply products of mathematical imagination but were originally proposed to describe phases of ferro and antiferroelectric materials. \cite{Slater} Thus they are effective descriptions of complex organization of matter. These vertex models are classical because they are made out of  commuting variables. 

The interest in the quantum six-vertex model, to be defined below, is rooted in the unusual phenomenology of the cuprate high temperature superconductors.~\cite{Chakravarty} It has been argued that the origin of the pseudogap is an unconventional broken symmetry in which a particle and a hole is bound in an angular momentum $l=2$ state, resulting in an order parameter, the $d$-density wave (DDW),  which is effectively hidden.~\cite{Chakravarty2} 
The statistical mechanical description of DDW, which includes both thermal and quantum fluctuations of the directions of the bond currents, is the quantum six-vertex model.~\cite{Chakravarty} In particular, the description of DDW in terms of  the six-vertex model resolves the puzzle as to why there are no associated specific heat anomalies, as   the phase transition corresponds to an essential singularity in the free energy. In this Letter we consider commensurate DDW as possible incommensuration wave vectors tend to be small in extended Hubbard models.\cite{ICDDW}

Among the observed broken symmetries in the particle-particle channel are $s$, $p$, and $d$-wave superconductors, where the Cooper pairs bind in the angular momentum channels $l=0, 1, \text{and}\; 2$ respectively. It is remarkable however, that in the particle-hole channel, the observed broken symmetries are mainly confined to $s$-wave symmetry: $s$-wave singlet and triplet density waves, known as the charge and spin density waves respectively.\cite{Nayak}

Here we investigate, for the first time, the phase diagram of the quantum six-vertex model, which is an interesting and unusual  statistical mechanical model in itself. A more practical  goal is to examine the stability of DDW with respect to order parameter fluctuations. Until now, DDW has been treated within mean field theory; see, however, Ref.~\cite{Chakravarty}. A full treatment of fermionic excitations is complex, but quenched fermions can be incorporated as sources and sinks by enlarging the model to the eight vertex model. We show that DDW is remarkably stable against quantum fluctuations. In addition,  we note an unsuspected connection between the quantum six-vertex model and the $(2+1)$-dimensional  $n$-vector ($n=2$) model with cubic anisotropy, which vindicates the numerical result that quantum fluctuations drive the DDW   to a new phase, termed the resonating plaquette phase, via a first order transition.  

\begin{figure}
\includegraphics[clip,width=8cm]{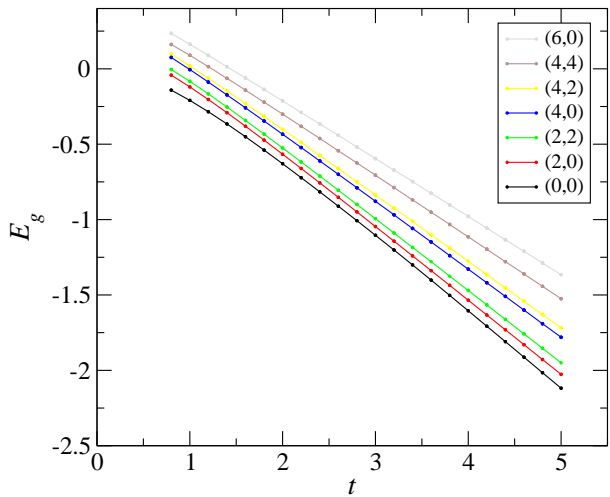}
\caption{(color online) Ground state energy per plaquette vs. $t$ for different topological sectors labelled by the total arrow polarization in the $x$ and $y$-directions. ($L=12$).
\label{EgTop}}
\end{figure}

Our quantum six-vertex model bears some resemblance to the widely discussed quantum dimer model,  but it does not  exhibit the Rokhsar-Kivelson (RK) point \cite{Rokhsar,Sondhi} and  differs from another recent attempt to quantize  vertex models. \cite{Fradkin} The latter, in which the RK point was built in by construction, had  very different goals: the ground state degeneracy was exploited to provide, in part, an understanding of topologically protected quantum computation. \cite{Kitaev,Freedman}  Although our model contains topological sectors, distinguished by their total arrow polarizations in  $x$ and $y$- directions, the lowest energy belongs to the trivial topological sector of DDW, with zero total polarization (on an even-even lattice), see Fig.~\ref{EgTop}. For other interesting models involving constrained quantum dynamics, see Refs.~\cite{Moore,Chamon}.

The Hamiltonian is
\be
H = \sum_{v \in {\rm vertices}} E(v)  -t \sum_{p \in {\rm plaq.}} \Big( | \clockwise \rangle \langle \counterclockwise | + | \counterclockwise \rangle \langle \clockwise |  \Big) 
\label{Hamiltonian}
\ee
A configuration in this model consists of a set of interconnected vertices that have four edges, arrows, which point either away from or into a vertex. In the six-vertex model only vertices with two arrows pointing into and two arrows pointing away from a vertex are allowed. The set of all possible vertex configurations forms the Hilbert space of the model.
The first term in  Eq.~(\ref{Hamiltonian}) assigs a potential energy to a vertex configuration. We will choose $E( \conevertex) = E(\ctwovertex)= 0$ and $E( \aonevertex ) =E(\atwovertex)=E(\bonevertex)=E(\btwovertex)=1$. The classical ground state is then DDW. The second term is quantum mechanical and reverses the arrows around elementary flippable plaquettes .  

We investigate our model using a continuous-time diffusion Monte Carlo simulation\cite{Olav} combined with the forward-walking technique\cite{forward} to extract ground state expectation values. The accuracy of the simulations were improved by including a guiding wave function proportional to a power, $\alpha$, of the number of flippable plaquettes in the evolved state. The value of $\alpha$ was determined by minimizing the variance of the local energy\cite{minvar} in short trial runs prior to the actual simulations. Simulation results for small systems with linear size $L=4$ were checked against exact diagonalization results, and the largest system studied had $L=20$.  

\begin{figure}
\includegraphics[clip,width=8cm]{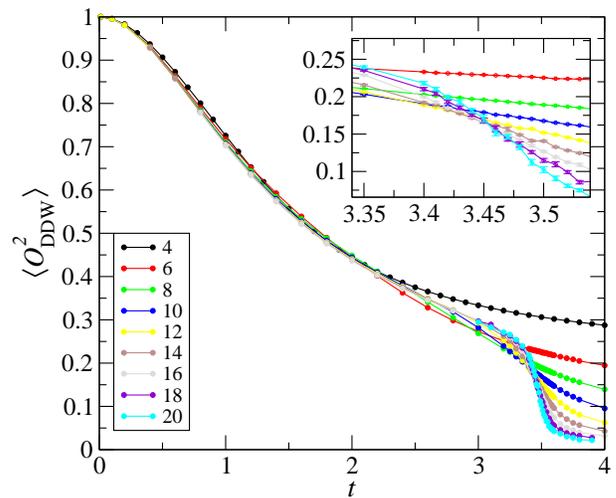}
\caption{(color online) The squared DDW order parameter vs. t for different lattice sizes. The inset is a blowup of the region near the phase transition.
\label{afe2}}
\end{figure}

The DDW order parameter is defined as $\DDW = \f{1}{4L^2} \sum_{p \in {\rm plaq}} (-1)^{p_x+p_y} C_p$, where $p_x$($p_y$) is the integer $x$($y$)-coordinate of the plaquette center $p$. $C_p$ sums the arrows around $p$. An arrow pointing in the counter-clockwise(clockwise) direction gives a contribution +1(-1). Fig.~\ref{afe2} shows $\langle \DDW^2 \rangle$ as a function of $t$ for different linear lattice sizes $L$.  With increasing $L$ the curves appear to develop into a discontinuity at $t=t_c \approx 3.47$, indicating a phase transition.

\begin{figure}
\includegraphics[clip,width=8cm]{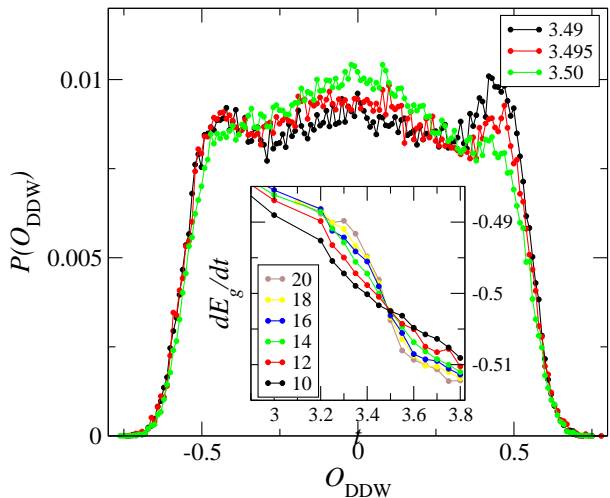}
\caption{(color online) Histograms of $\DDW$ at $L=20$ for different values of $t$ close to the transition. The inset shows the numerically differentiated (finite-difference) ground state energy with respect to $t$ for different system sizes.
\label{hist20}}
\end{figure}

To investigate the nature of this phase transition we construct histograms of $\DDW$ consistent with the forward-walking method. The appropriate bin to increment is selected from the value of $\DDW$ occuring at a long time $\Delta \tau=3$ in the past. A weight that corrects for the population bias needed to control the simulation is then added to this bin. The histogram is finally normalized by the total sum of weights.  
For small $t$ the histograms are peaked at large positive and negative values that move towards zero as $t$ is increased. However this motion is not continuous. For $t$ close to $t_c$ a 3-peak structure indicating a coexistence of phases\cite{Challa} is visible for the largest system sizes, Fig.~\ref{hist20}. When $t$ is tuned through $t_c$ the peaks do not move, rather their relative weights change, indicating a discontinuous change in $\DDW$ in the thermodynamic limit.  That the transition is first order is also indicated by measuring the differentiated ground state energy per plaquette with respect to $t$. This results in numerical curves that develop into a discontinuity at the transition as the system size is increased, see the inset of Fig.~\ref{hist20}.
   
\begin{figure}
\includegraphics[clip,width=8cm]{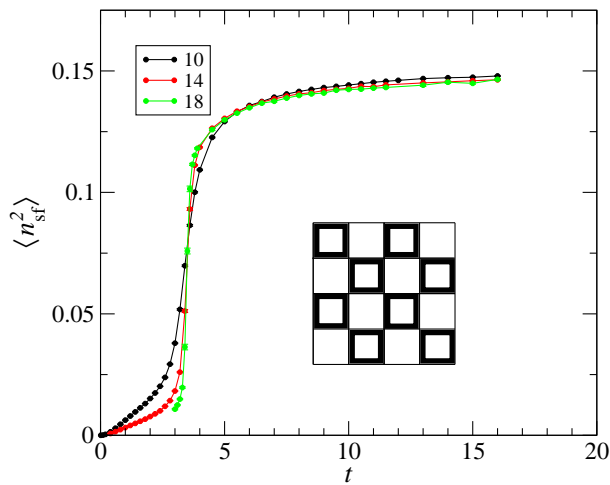}
\caption{(color online) The density of staggered flippable plaquettes squared vs. $t$ for different system sizes. The figure indicates the resonating plaquette phase. Clockwise and counterclockwise arrangements of arrows are resonating on the ``fat'' plaquettes.
\label{sflip}}
\end{figure}

What is the ground state for $t$ larger than  $t_c$? To answer this question  we measure the density of flippable plaquettes $\nf= (1/L^2) \sum_{p} F_p$. $F_p$ is 1 if the plaquette $p$ is flippable, and is $0$ otherwise. We find that $\nf$ drops fast from 1 and saturates to a value $\nf \approx 0.56$ for large values of $t$ independent of system size. Thus the large-$t$ ground state contains a significant number of flippable plaquettes. Another interesting quantity is the {\em staggered} density of flippable plaquettes $\nsf = (1/L^2) \sum_p (-1)^{p_x+p_y} F_p$. A plot of $\langle \nsftwo \rangle$ vs. $t$ for different system sizes is shown in Fig.~\ref{sflip}.
The staggered density is 0 below the transition and finite above. Thus there is a preference for flippable plaquettes on one of the two sublattices. At high values of $t$, $|\nsf |\approx 0.39$. For $t>t_c$ circulating arrows around plaquettes on one of the sublattices ``resonate'', changing their circulation direction constantly. We term this a resonating plaquette phase. This phase occurs also in a quantum six-vertex model with a built-in RK-point\cite{Shannon} and in the Heisenberg model on a planar pyrochlore lattice.\cite{Fouet} 

\begin{figure}
\includegraphics[clip,width=8cm]{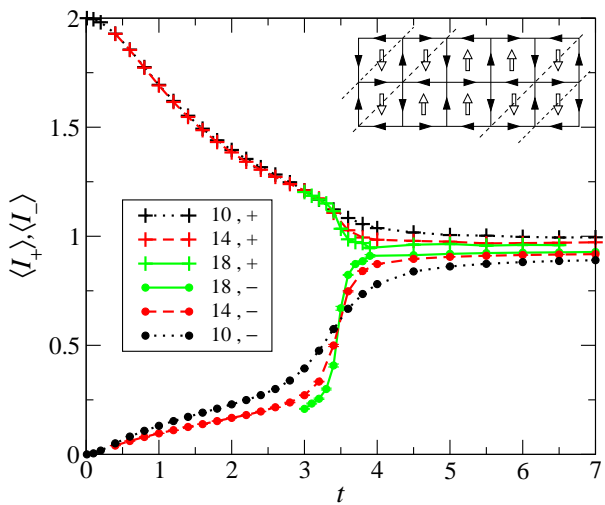}
\caption{(color online) $\langle I_- \rangle$ (pluses) and $\langle I_+ \rangle$ (circles) vs. $t$ for different system sizes. Upper right corner: DDW configuration in terms of Ising spins.
\label{ising}}
\end{figure}

In order to identify proper order parameters for the quantum six-vertex model and infer the associated finite temperature phase diagram we write
a six-vertex configuration in terms of Ising spins at the center of the plaquettes. A vertex arrow points in the positive (negative) coordinate direction if its neighboring Ising spins are equal (opposite), see Fig.~\ref{ising} upper right corner. By performing a $\pi$-rotation about the spin $x$-axis for spins on the dashed diagonal lines in Fig.~\ref{ising}, the DDW shown can be identified with the ferromagnetic state. In this new basis the Hamiltonian for the quantum six-vertex model becomes that of two inter-penetrating ferromagnetic Ising models, one on each sublattice, coupled by a frustrated four-spin term that couple Ising spins centered around a vertex. The couplings are chosen such that sources and sinks in the arrow-configurations are penalized energetically and a transverse field that flips the Ising spins plays the role of the quantum term. We replace the Ising spins with continuum fields $S_i$, the label indicating the sublattice, and a term proportional to $(S_i^2-1)^2$ to preserve the spin magnitude. Thus,  one arrives at the continuum action of an $n$-vector $(n=2)$ model with cubic anisotropy, neglecting irrelevant higher order derivatives :
\bea
	S & = & \int_0^{\beta c} dx_0 \int d^{2} x  \left(
	\f{1}{2} \sum_i  \left( \f{\partial S_i}{\partial x_0} \right)^2 
	+\f{1}{2} \sum_i  \left( \f{\partial S_i}{\partial \vec{x}} \right)^2  \right. \nonumber \\
	&  & \left. + m\sum_i S_i^2 + u (\sum_i S_i^2)^2 + v \sum_i S_i^4 \right). \label{nvectoraction}
\eea

For our purposes, namely identifying the order parameters,
the precise values of the  coefficients are not necessary. The isotropic $n=2$ vector model with cubic anisotropy has been extensively studied.\cite{nvector}  It has two ordered states for $m<0$. For $v>0$ it orders diagonally. That is both Ising models (on both sublattices) have identical magnitudes of magnetization. For $v<0$ the $n$-vector model exhibits axis ordering: only one of the Ising models orders and the other remains disordered.  The transition between these two ordered phases is first order.  We identify the quantum phase transition in the quantum six-vertex model with this transition, and infer that the proper order parameters are the combinations of Ising order parameters on the two sublattices.

We measure these order parameters in a simulation of Eq.~\ref{Hamiltonian} by assigning the value $\sigma=+1$ to all Ising spins in the starting DDW-state. During the evolution a spin is flipped if the corresponding currents around the plaquette is reversed, otherwise it is unchanged.  The Ising order parameter on sublattice $i$ is $I_i=(2/L^2) \sum_p \sigma_p$ and takes values between $-1$ and $1$. In Fig.~\ref{ising} we have plotted $I_\pm=| I_1^2 \pm I_2^2 |$ as functions of $t$. For $t<t_c$ $I_+$ is finite and $I_-$ approaches zero for large system sizes indicating identical magnetization magnitude on the two sublattices. For $t>t_c$, both $I_+$ and $I_-$ approaches the same finite value ($L \to \infty$), meaning that only one sublattice orders.
The single-sublattice Ising order is an order by disorder phenomenon, as it is stabilized by the frequent plaquette flips, the resonating plaquettes, on the other sublattice. This makes the occurence of flippable plaquettes on the ordered sublattice infrequent. 

\begin{figure}
\includegraphics[clip,width=8cm]{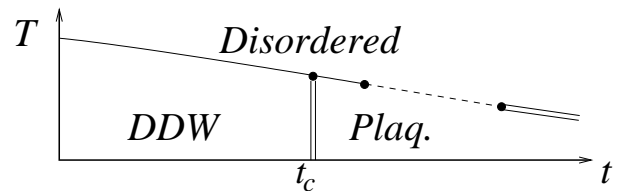}
\caption{Possible phase diagram for the quantum six-vertex model as functions of flipping strength $t$ and temperature $T$ showing first order transitions (double lines), KT transitions (solid lines) and continuous transitions with varying exponents (dashed line).
\label{phasediagram}}
\end{figure}

 As the temperature is increased, the action (\ref{nvectoraction}) will describe a two-dimensional $n$-vector model with cubic anisotropy, with renormalized values of $u$ and $v$. For $v=0$ the transition is governed by an XY-fixed point, implying  a Kosterlitz-Thouless (KT) transition above $t_c$ on the boundary between the ordered phases in the quantum six-vertex model. We will assume this also holds for small positive values of $v$ and $u$ in order to be consistent with the known fact that the $t=0$ thermal transition is KT-like.\cite{Kadanoff} It then follows by continuity that the thermal disordering of the DDW phase is in the KT universality class for all $0<t<t_c$. For $t>t_c$ ($v<0$) we also expect a KT transition in the vicinity of $t_c$ assuming that the renormalization group (RG) flows are still governed by the XY-point; however it is possible that the system will follow RG flows directed towards the fixed line existing between the XY and the Cubic fixed point\cite{Calabrese} and that the KT-transition will change into a line of transitions with continuously varying exponents. On further increasing $t$ this line might terminate and become first order consistent with the fluctuation-driven first order transitions known to exist in the n-vector model.\cite{nvector} We have sketched this possiblity in Fig.~\ref{phasediagram}. A direct transition from KT-like behavior to first order transitions is also consistent with the RG flows for the $n$-vector model.

The vertex models appear to be complex, but they are capable of encompassing unconventional order parameters such as orbital antiferromagnets, spin nematics, excitonic condensates, ice models, flux phases, and the possible DDW state in the cuprates. Their quantum versions hold many surprises  and are likely to exhibit fractionalized excitations and non-trivial topological effects. In the present paper we have shown that they shed considerable light on the stability of the DDW against a class of quantum fluctuations and lead to a novel quantum state of matter, the resonating plaquette state.

S. C. acknowledges a grant from the National Science Foundation: NSF-DMR 0411931.

 \end{document}